\documentstyle[aps,twocolumn,pra,epsf,eqsecnum]{revtex}
\begin{document}

\title{Anharmonic parametric excitation in optical lattices}

\author{R. J\'auregui}
\address{Instituto de F\'{\i}sica, Universidad Nacional Aut\'onoma de
M\'exico, Apdo. Postal 20-364, M\'exico, 01000, D.F., M\'exico}
\author {N. Poli, G. Roati, and G. Modugno}
\address{INFM-European Laboratory for Nonlinear Spectroscopy (LENS),
Universit\`a di Firenze, Largo E. Fermi 2, 50125 Firenze, Italy}

\date{\today}
\draft
\maketitle

\begin{abstract}
  We study both experimentally and theoretically the losses induced by
  parametric excitation in far--off--resonance optical lattices. The
  atoms confined in a 1D sinusoidal lattice present an excitation
  spectrum and dynamics substantially different from those expected
  for a harmonic potential. We develop a model based on the actual
  atomic Hamiltonian in the lattice and we introduce semiempirically a
  broadening of the width of lattice energy bands which can physically
  arise from inhomogeneities and fluctuations of the lattice, and
  also from atomic collisions. The
  position and strength of the parametric resonances and the evolution
  of the number of trapped atoms are satisfactorily described by our
  model.
\end{abstract}

\section{Introduction}
The phenomenon of parametric excitation of the motion of cold trapped
atoms has recently been the subject of several theoretical and
experimental investigations \cite{savard97,friebel98,gardiner2000}.
The excitation caused by resonant amplitude noise has been proposed as
one of the major sources of heating in far--off--resonance optical
traps (FORTs), where the heating due to spontaneous scattering forces
is strongly reduced\cite{miller93}. In particular, the effect of
resonant excitation is expected to be particularly important in
optical lattices, which usually provide a very strong confinement to
the atoms, resulting in a large vibrational frequency and in a
correspondingly large transfer of energy from the noise field to the
atoms\cite{savard97}.\par Nevertheless, parametric excitation is not
only a source of heating, but it also represents a very useful tool to
characterize the spring constant of a FORT or in general of a trap for
cold particles, and to study the dynamics of the trapped gas.  Indeed,
the trap frequencies can be measured by intentionally exciting the
trap vibrational modes with a small modulation of the amplitude of the
trapping potential, which results in heating \cite{vuletic98} or
losses\cite{friebel98,roati01} for the trapped atoms when the
modulation frequency is tuned to twice the oscillation frequency. This
procedure usually yields frequencies that satisfactorily agree with
calculated values, and are indeed expected to be accurate for the
atoms at the bottom of the trapping potential.  From the measured trap
frequencies is then possible to estimate quantities such as the trap
depth and the number and phase space densities of trapped atoms. We
note that this kind of measurement is particularly important in
optical lattices, since the spatial resolution of standard imaging
techniques is usually not enough to estimate the atomic density from a
measurement of the volume of a single lattice site.\par Recently, 1D
lattices have proved to be the proper environment to study collisional
processes in large and dense samples of cold atoms, using a trapping
potential independent for the magnetic state of the atoms. In this
systems, the parametric excitation of the energetic vibrational mode
along the lattice provides an efficient way to investigate the
cross-dimensional rethermalization dynamics mediated by elastic
collisions \cite{vuletic99,roati01}.\par

Most theoretical studies of parametric excitation rely on a classical
\cite{landau} or quantum \cite{savard97} harmonic approximation of the
confining potential. Under certain circumstances these expressions
show quite good agreement with experimental results
\cite{gardiner2000}. However, general features of the optical lattice
could be lost in these approaches. For example, a sinusoidal potential
exhibits an energy band structure and a spread of transition
energies, while harmonic oscillators have just a discrete equidistant
spectrum. Thus, we might expect that the excitation process may happen
at several frequencies, and with a non-negligible bandwidth. Such
anharmonic effects can be important whenever the atoms are occupying a
relatively large fraction of the lattice energy levels. The purpose of
this paper is to give a simple description of parametric excitation in
a sinusoidal 1D lattice.  In the next Section, we briefly discuss
general features of the stationary states on such a lattice.  Then, we
summarize the harmonic description given in Ref.\cite{savard97} and
extend it to the anharmonic case. By a numerical evaluation of
transition rates, we make a temporal description of parametric
excitation which is compared with experimental results. We discuss
about the relevance of broadening of the spectral lines to understand
the excitation process in this kind of systems. Some conclusions are
given in the last Section.

\section{Stationary states of a sinusoidal optical lattice.}

The Hamiltonian for an atom in a red detuned FORT is
\begin{equation}
H = \frac{ P^2}{2M} +V_{\rm eff}(\vec x)\,, \label{h:1}
\end{equation}
with
\begin{equation}
V_{\rm eff}(\vec x) =  -\frac{1}{4}\alpha \vert {\cal E}(\vec x)\vert
^2\,,
\end{equation}
where $\alpha$ is the effective atomic polarizations and ${\cal E}(x)$
is the radiation field amplitude. For the axial motion in a
sinusoidal 1D lattice we can take
\begin{eqnarray}
H_{ax} &=& \frac{P^2_z}{2M}
+ V_0 \cos^2 (k z)\\
&=& \frac{P^2_z}{2M} +\frac{V_0}{2}\big( 1 + \cos (2kz))\big).
\label{h:s}
\end{eqnarray}
The corresponding stationary Schr\"odinger equation
\begin{equation}
-\frac{\hbar^2}{2M}\frac{d^2\Phi}{dz^2} + \frac{V_0}{2} \big (1 +
\cos (2kz) \big) \Phi = E \Phi
\end{equation}
can be written in canonical Mathieu's form
\begin{equation}
\frac{d^2\Phi}{du^2} + (a - 2q \cos 2u)\Phi = 0 \label{mathieu:1}
\end{equation}
with
\begin{equation}
a = \big(E- \frac{V_0}{2}\big)\big(\frac{2M}{\hbar^2 k^2}\big)
\quad 2q = \frac{V_0}{2} \big(\frac{2M}{\hbar^2 k^2}\big).
\end{equation}
It is well known that there exists countably infinite sets of
characteristic values $\{a_r\}$ and $\{b_r\}$ which respectively yield
even and odd periodic solutions of Mathieu equation. These values
also separate regions of stability.  In particular, for $q\ge 0$ the
band structure of the sinusoidal lattice corresponds to energy
eigenvalues between $a_r$ and $b_{r+1}$ \cite{abramowitz}. The
unstable regions are between $b_r$ and $a_r$. For $q>>1$, there is an
analytical expression for the band width \cite{abramowitz}:
\begin{equation}
b_{r+1} - a_r \sim 2^{4r+5} \sqrt{2/\pi}q^{\frac{1}{2}r
+\frac{3}{4}} e^{-4\sqrt{q}}/r! . \label{width:1}
\end{equation}
The quantities defined above can be expressed in terms of a frequency
$\omega_0$ defined in the harmonic approximation of the potential 

\begin{equation} \frac{1}{2} M \omega_0^2 =
\frac{V_0}{2}\frac{(2k)^2}{2!}\,,
\label{harm}
\end{equation}
thus obtaining
\begin{equation}
a = \big(E-\frac{V_0}{2}\big)\big(\frac{4V_0}{\hbar^2
\omega_0^2}\big) ; \quad\quad q = \big( \frac{V_0}{\hbar \omega_0}
\big)^2\,.
\end{equation}
Thus, the width of the r-band can be estimated using
Eq.~(\ref{width:1}) whenever the condition $(V_0/\hbar\omega_0)^2 >>1$
is satisfied. In the experiment we shall be working with a 1D optical
lattice having $V_0 \sim 10.5\hbar \omega_0$. While the lowest
band $r=0$ has a negligible width $\sim 10^{-18} \hbar \omega_0$, the
band widths for highest lying levels $r=10,11,12$,and $13$ would
respectively be $0.0065$ ,$0.1036$, $1.52$, and $20.56$ in units of
$\hbar\omega_0$.

In order to determine the energy spectrum, a variational calculation
can be performed. We considered a harmonic oscillator basis set
centered in a given site of the lattice, and with frequency
$\omega_0$. The diagonalization of the Hamiltonian matrix associated
to (\ref{h:s}) using 40 basis functions gives the eigenvalues
$E_n<V_0$ shown in Table I for $V_0=10.5\hbar\omega_0$. According to
the results of last paragraph, the eigenvalues $12$ and $13$ belong to
the same band while the band width for lower levels is smaller than
$0.11 \hbar\omega_0$

\section{Parametric excitation}
As already mentioned, parametric excitation of the trapped atoms consists
in applying a small modulation to the intensity of the trapping
light,
  \begin{equation}
 H = \frac{P^2}{2M} + V_{\rm eff}[1+\epsilon(t)].
 \label{general:1}
 \end{equation}
 Within first order perturbation theory, this additional field
 induces transitions between the stationary states $n$ and $m$
 with an averaged rate
 \begin{eqnarray}
 R_{m\leftarrow n} &=& \frac{1}{T} \left| \frac{-i}{\hbar} \int_0^T
 dt  T(m,n) \epsilon(t)
 e^{i\omega_{mn}t}\right|^2\nonumber\\
 &=& \frac{\pi}{2\hbar^2}\vert T(m,n)\vert^2 S(\omega_{mn});
 \quad \omega_{mn} = \frac{E_m-E_n}{\hbar}
 \label{Rmn:1}
 \end{eqnarray}
 where
 \begin{eqnarray}
 T(m,n)&=&\langle m\vert V_{\rm eff}\vert n\rangle\nonumber\\
 &=& E_n \delta_{nm} - \frac{1}{2M}\langle m\vert \hat P^2\vert
 n\rangle
 \end{eqnarray}
 is the matrix element of the space part of the perturbation and
 \begin{equation}
 S(\omega) =  \frac{2}{\pi} \int_0^T d\tau \cos \omega\tau
 \langle \epsilon(t) \epsilon(t+\tau)\rangle
 \end{equation}
 is the one-sided power spectrum of the two-time correlation
 function associated to the excitation field amplitude.

 If the confining potential is approximated by a harmonic well,
 the transition rates different from zero are
 \begin{eqnarray}
 R_{n\leftarrow n} & = & \frac {\pi\omega_0^2}{16} S(0)(2n+1) \\
 R_{n\pm 2\leftarrow n} &=& \frac{\pi\omega_0^2}{16}S(2\omega_0)
 (n+1\pm 1)(n\pm 1)
 \end{eqnarray}
 The latter equation was used in \cite{savard97} to
 obtain a simple expression for the heating rate,
 \begin{equation}
 \langle \dot E \rangle = \frac{\pi}{2}\omega_{0}^2 S
 (2\omega_0) \langle E \rangle\,, \label{hexp:1}
 \end{equation}
 showing its exponential character. The dependence on $2\omega_0$
 is characteristic of the parametric nature of the excitation process.
 The fact that $\hbar$ is not present is consistent with
 the applicability of Eq.~(\ref{hexp:1}) in the classical regime.
 
 Classically, parametric harmonic oscillators exhibit resonances not
 just at $2\omega_0$ but also at $2 \omega_0/n$ with $n$ any natural
 number \cite{landau}. In fact, the resonances corresponding to $n$=2,
 {\em i.e.} at an excitation frequency $\omega_0$, have been observed
 in optical lattices \cite{friebel98,roati01}.  A quantum description
 of parametric harmonic excitation also predicts resonances at the
 same frequencies via $n$-th order perturbation theory
 \cite{jauregui01}.  In particular, the presence of the resonance at
 $\omega_0$ can be justified with the following argument. According to the
 standard procedure, the second order correction to the transition
 amplitude between states $\vert n\rangle$ and $\vert m\rangle$ is
 given by
\begin{eqnarray}
 R^{(2)}_{m\leftarrow n}=\langle n \vert {\cal U}^{(2)}(t_0,t)\vert m \rangle = 
\sum_{k}
\big(\frac{-i}{\hbar}\big)^2 T(n,k)T(k,m)\nonumber\\
\int_{t_0}^t dt^\prime e^{i\omega_{nk}t^\prime} \epsilon(t^\prime) \int_{t_0}^t
dt^{\prime\prime} e^{i\omega_{km}t^{\prime\prime}}
\epsilon(t^{\prime\prime}) \label {2step}
\end{eqnarray}
 with ${\cal U}^{(2)}(t_0,t)$ the second order correction to the
 evolution operator ${\cal U}$. Therefore, the transition may be
 described as a two step procedure $\vert m\rangle \leftarrow\vert k
\rangle\leftarrow \vert n\rangle$.  For harmonic parametric excitation
 the matrix element of the space part of the perturbation differs from
 zero just for transitions $\vert n\rangle \leftarrow \vert n\rangle$
 and $\vert n\pm 2\rangle \leftarrow \vert n\rangle$. Consider a
 transition in Eq.~(\ref{2step}) involving a "first" step in which the
 state does not change $\vert n\rangle \leftarrow \vert n \rangle$ and
 a "second" step for which $\vert n\pm 2 \rangle \leftarrow \vert n
 \rangle$. Then resonance phenomena occur when the total energy of the
 two excitations, 2$\hbar\Omega$, coincides with that of the second step
 transition, $i.$ $e.$ for an excitation frequency $\Omega=\omega_o$.
 
 These ideas can be directly extended to anharmonic potentials: the
 corresponding transition probability rates $R(n,m)$ would be
 determined by the transition matrix $T(n,m)$, by the transition
 frequencies $\omega_{nm}$ and by the time dependence of the
 excitation $\epsilon(t)$.  In general, anharmonic transition matrix
 elements $T(n,m)$ will be different from zero for a wider set of
 pairs $(n,m)$. Besides, the transition energies will not be unique so
 that the excitation process is $not$ determined by the excitation
 power spectrum at a single given frequency $2\omega_0$ and its
 subharmonics $2\omega_o/n$. As an example the transition energies for
 the specific potential considered in this work are reported in Table~I. 
Therefore, within the model Hamiltonian of Eq.~(\ref{general:1}),
 resonance effects can occur for several frequencies that may alter
 the shape of the population distribution within the trap. However, in
 general these resonant excitations will not be associated with the
 escape of trapped atoms.

 Here we are interested in a 1D lattice; the direct
 extension of the formalism mentioned above requires the evaluation of
 the matrix elements $T(n,m)$ among the different Mathieu states that
 conform a band. This involves integrals which, to our knowledge, lack
 an analytical expression and require numerical evaluation. As an
 alternative, we consider functions which variationally approximate the
 Mathieu functions. They are the eigenstates of the Hamiltonian
 (\ref{h:s}) in a harmonic basis set of frequency $\omega_0$:
 \begin{equation}
 \vert n\rangle = \sum_{i=1}^{i_{max}} c_{ni} \vert
 i\rangle_{\omega_0},
 \label{levels}
 \end{equation}
 These states are ordered according to their energy: $E_n\le E_{n+1}$
 as exemplified in Table I.
 Within this scheme one obtains a very simple expression
 for $T(n,m)$
 \begin{equation}
 T(n,m) = E_n \delta_{nm}
 - \sum_{i,j=1}^{i_{max}} c_{ni} c_{nj}\frac{1}{2M}\langle
 i\vert \hat P^2\vert j \rangle.
 \end{equation}
 It is recognized that any discrete basis set approximation to a
 system with a band spectrum will lack features of the original
 problem which have to be carefully analyzed. Anyway, alternatives to
 a discrete basis approach may be cumbersome and not necessarily yield
 a better approach to understand general properties of experimental
 data. While the discrete basis approach is exact for transitions
 between the lowest levels, which have a negligible width, eigenstates
 belonging to a band of measurably width should be treated with
 special care. Thus, we shall assume that matrix elements $T(\nu,\mu)$
 involving states with energies ${\cal E}_\nu$ and ${\cal E}_\mu$, so
 that $E_n-(E_n - E_{n-1})/2\le{\cal E}_\nu \le E_n+(E_{n+1}-E_n)/2$
 with an analogous expressions for ${\cal E}_\mu$, are well
 approximated by $T(n,m)$.

 Within this scheme the equations which describe the probability
 $P(n)$ of finding an atom in level $n$, given the transition rates
 $R_{m\leftarrow n}$ are
 \begin{equation}
 \dot P(n) = \sum_{m} R^{(1)}_{m\leftarrow n} (P(m) - P(n))
 \label{difeq}
 \end{equation}
 in the first order perturbation theory scheme, and the
 finite difference equations
\begin{eqnarray}
P_n(t)&=&P_n(t_0)+\nonumber\\& & \sum_m R^{(1)}_{m\leftarrow n}(P_m(t_0)-
P_n(t_0))(t-t_0)+
\nonumber\\
& & \sum_m R^{(2)}_{m\leftarrow n}(P_m(t_0)-P_n(t_0)) (t-t_0)^2\,,
\label{fdiff}
\end{eqnarray}
 valid up to second order time-dependent perturbation theory
 whenever $t\sim t_0$. Both sets of equations are subjected to the condition
 \begin{equation}
 \sum_{n}P(n) = 1.
 \end{equation}
 
 Now, according to Eqs.~(\ref{Rmn:1}) and Eq~(\ref{2step}), the
 evaluation of $R^{(r)}_{n\leftarrow m}$ also requires the
 specification of the spectral density $S(\omega)$.  In the problem
 under consideration, the discrete labels $m,n$ are used to calculate
 interband transitions which are actually spectrally broad. This
 broadening might arise not only from the band structure of the energy
 spectra associated with the Hamiltonian Eq.~(\ref{h:s}), but also
 from other sources, which we will discuss below. Broad spectral lines
 can be introduced in our formalism by defining an effective spectral
 density $S_{\rm eff}(\omega)$ which should incorporate essential
 features of this broadening without simulating specific
 features. Having this in mind, an effective Gaussian density of
 states ${\cal S}_n(\omega)$ is associated to each level $\vert
 n\rangle$ of energy $E_n$
 \begin{equation}
 {\cal S}_n(\omega) = \frac{1}{\sqrt{2\pi}\sigma_n}\,e^{-\displaystyle
 \frac{(\hbar\omega -E_n)^2}{2(\hbar\sigma_n)^2}}.
 \label{S:discrete}
  \end{equation}
  The spectral effective density $S_{\rm eff}(\omega_{nm})$ associated
  to the transition $m\leftarrow n$ is obtained by the convolution of
  ${\cal S}_n(\omega)$ with ${\cal S}_m(\omega)$ and with the
  excitation source spectral density $S(\omega)$. For a monochromatic
  source the latter is also taken as a Gaussian centered at the
  modulation frequency that once integrated over all frequencies
  yields the square of the intensity of the modulation source.  The
  net result is that $S_{\rm eff}(\omega_{nm})$ has the form
  \begin{equation}
  S_{\rm eff}(\omega) = S_0\,e^{-\displaystyle\frac{(\omega-\omega_{\rm 
eff})^2}{2\sigma_{\rm eff}^2}}
\end{equation}
 with $\omega_{\rm eff}$  determined by the modulation frequency
 $\Omega$ and the energies $E_n$ and $E_m$. The effective width
 $\sigma_{\rm eff}$ contains information about the frequency
 widths of the excitation source and those of each level.

\section{Comparison with experimental results.}
 We have tested the procedure described in last section to model
 parametric excitation in a specific experiment conducted at LENS.  In
 this experiment $^{40}$K fermionic atoms are trapped in a 1D lattice,
 realized retroreflecting linearly polarized light obtained from a
 single--mode Ti:Sa laser at $\lambda$=787\,nm, detuned on the red of
 both the D$_1$ and D$_2$ transition of potassium, respectively at
 769.9\,nm, and 766.7\,nm. The laser radiation propagates along the
 vertical direction, to provide a strong confinement against
 gravity. The laser beam is weakly focused within a two-lens telescope
 to a waist size $w_0\simeq 90\,\mu$m, with a Rayleigh length
 $z_R$=3\,cm; the effective running power at the waist position is
 $P$=350\,mW.
 
\begin{figure}
\begin{center}
\leavevmode\epsfxsize=8cm
    \epsfbox{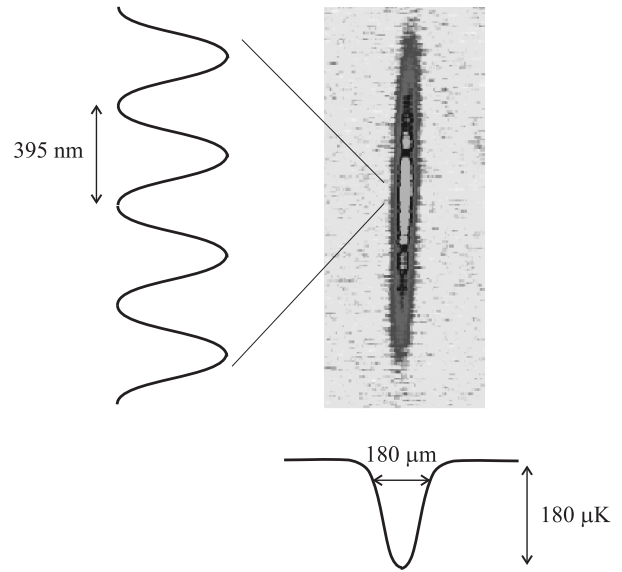}
\end{center}
\caption{Absorption image of the atoms in the optical lattice, and
    shape of the optical potential in the two relevant directions.}
\label{fig1}
\end{figure}

 The trap is loaded from a magneto-optical trap (MOT), thanks to a
 compression procedure already described in \cite{roati01}, with about
 $5 \times 10^5$ atoms at a density around $10^{11}\,$cm$^{-3}$. The
 typical vertical extension of the trapped atomic cloud, as detected
 with a CCD camera (see Fig.~\ref{fig1}), is $500\,\mu$m,
 corresponding to about 1200 occupied lattice sites with an average of
 400 atoms in each site. Since the axial extension of the atomic cloud
 is much smaller than $z_R$, we can approximate the trap potential to
\begin{equation}
V(r,z)=V_0\, e^{\displaystyle-\frac{2r^2}{w_0^2}}\,\cos^2(kz)\,;\quad
k=2\pi/\lambda\,,
\end{equation}
 thus neglecting a 5\% variation of $V_0$ along the lattice. The atomic
 temperature in the lattice direction is measured with a
 time--of--flight technique and it is about 50\,$\mu$K.
 
\begin{figure}
\begin{center}
\leavevmode\epsfxsize=9cm
    \epsfbox{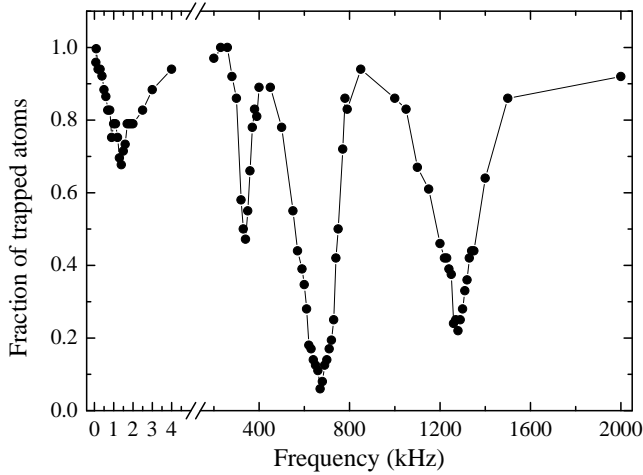}
\end{center}
\caption{Experimental spectrum of the losses associated to parametric
    excitation of the trap vibrational modes. For the low and high frequency
    regions two different modulation amplitudes of 20\% and 3\%
    respectively, were used.}
\label{fig2}
\end{figure}

 In order to parametrically excite the atoms we modulate the intensity
 of the confining laser with a fast AOM for a time interval
 T$\simeq$100\,ms, with a sine of amplitude $\epsilon$=3\% and
 frequency $\Omega$. The variation of the number of trapped atoms is
 measured by illuminating the atoms with the MOT beams and collecting
 the resulting fluorescence on a photomultiplier. In Fig.\ref{fig2}
 the fraction of atoms left in the trap after the parametric
 excitation is reported {\em vs} the modulation frequency $\Omega/2\pi$.
 Three resonances in the trap losses are clearly seen at modulation
 frequencies 340\,kHz, 670\,kHz and 1280\,kHz. By identifying the
 first two resonances with the lattice vibrational frequency and its
 first harmonic, respectively, we get as first estimate $\omega_0
 \simeq 2\pi\times 340\,$kHz. As we will show in the following, these
 resonance are actually on the {\em red} of $\omega_0$ and
 2$\omega_0$, respectively, and therefore a better estimate is
 $\omega_0 \simeq 2\pi\times 360\,$kHz. Therefore the effective trap
 depth is, from Eq.~(9), $V_0 \simeq 185\,\mu$K $\simeq
 10.5\,\hbar\omega_0$. Since the atomic temperature is about
 $V_0$/3.5, we expect that most of the energy levels of the lattice
 have a nonnegligible population and therefore the anharmonicity of
 the potential could play an important role in the dynamics of
 parametric excitation. Note that the third resonance at high
 frequency, close to 4$\omega_0$, is not predicted from the harmonic
 theory. It is possible to observe also a much weaker resonance in the
 trap losses around 1.5\,kHz, which we interpret to be twice the
 oscillation frequency in the loosely confined radial direction.
 Anyway, in the following we will focus our attention just on the
 axial resonances.\par
 
 As discussed in the previous Section, the overall width of the
 excitation assumed for our model system could play an important role
 in reproducing essential features of experimental data. Since the
 source used in the experiment has a negligible line width, it is
 necessary to model just the broadening of the atomic resonances. The
 spread of the transition energies due to the axial anharmonicity is
 reported in Table I, while the broadening of each energy level, due
 to the periodic character of the sine potential, is estimated using
 Eq.~(\ref{width:1}). We now note that the 1D motion assumed in
 Section II is not completely valid in our case, since the atoms move
 radially along a Gaussian potential. Since the period of the radial
 motion is about 500 times longer than the axial period, the atoms see
 an effective axial frequency which varies with their radial position,
 resulting in a broadening of the transition frequency. Other sources
 of broadening are fluctuations of the laser intensity and pointing,
 and inhomogeneities along the lattice. We note that also elastic
 collisions within the trapped sample, which tend to keep a thermal
 distribution of the trap levels population, can contribute to an
 overall broadening of the loss resonances. Since it is not easy to build a
 model which involve all these sources, we introduce semiempirically
 an effective broadening for the $r$-th level (see
 Eq.~(\ref{S:discrete})).  Recognizing that the width could be energy
 dependent we considered the simple expression
\begin{equation}
\sigma_{r}^2 = \lambda_1 \big(\frac{E_r}{V_0}\big)^p +\lambda_0
\label{sigma:r}
\end{equation}
for several values of the constants $\lambda_1, \lambda_0$ and $p$.
When $p=0$, i.e. for a constant value of the band width we were not
able to reproduce the general experimental behavior reported in
Fig.\ref{fig2}. The best agreement between the simulation and the
experimental observations is obtained for $\lambda_0 =0.0002$,
$\lambda_1=0.0135$, in units of $\omega_0^2$, and $p=3$.  Similar
results are obtained also for slightly higher (lower) values of
$\lambda_{0,1}$ together with slightly higher (lower) values of the
power $p$. In Table I the resulting widths are shown for the lower
twelve levels. Note that we have intentionally excluded levels 11, 12 and
13 from the calculation, since their intrinsic width is so large that
the atoms can tunnel out of the trap along the lattice in much less
than 100\,ms \cite{tunnel}. Anyway, the inclusion of these levels
proved not to change substantially the result of the simulation.\par
 
\begin{figure}
\begin{center}
\leavevmode\epsfxsize=9cm
    \epsfbox{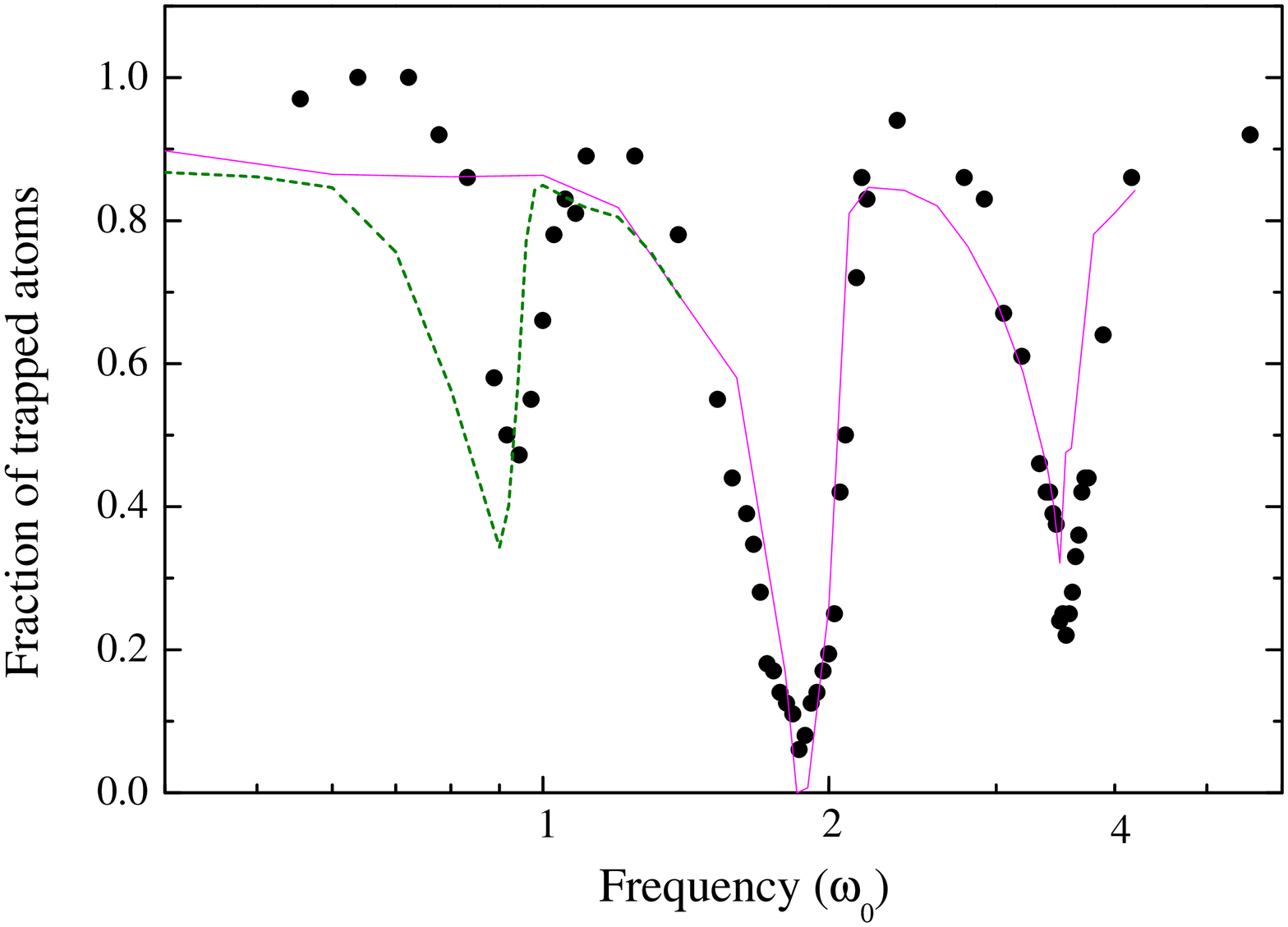}
\end{center}
\caption{
  Experimental (circles) and theoretical (lines) fraction of atoms
  left in the trap after parametric excitation {\em vs} the
  modulation frequency. The continuous line corresponds to the
  numerical integration of the first order perturbation theory
  equations (\ref{difeq}) and the dashed line to the numerical
  integration of the finite difference second order perturbation
  theory equations (\ref{2step}).} \label{fig3}
\end{figure}

The comparison of experimental and theoretical results is made in
Fig.~\ref{fig3}; the abscissa for the experimental data has been
normalized by identifying $\omega_0$ with $2\pi\times 360\,$kHz. As
already anticipated, the principal resonance in trap losses appears at
$\Omega\simeq 1.85\,\omega_0$. This result follows from the fact that
the excitation of the lowest trap levels is not resulting in a loss of
atoms, as it would happen for a harmonic potential. On the contrary,
the most energetic atoms, which have a vibrational frequency smaller
than the harmonic one, are easily excited out of the trap. The
asymmetry of the resonances, which has been observed also in
\cite{friebel98}, is well reproduced in the calculations and it is a
further evidence of the spread of the vibrational frequencies. The
first interesting result obtained by our study of parametric
excitation is therefore the correction necessary to extract the actual
harmonic frequency from the loss spectrum. For the specific conditions
of the present experiment, we find indeed that the principal resonance
in the trap losses appears at $\Omega\simeq 1.85\,\omega_0$. Anyway,
the calculation shows that the resonance is nearby this position for
all the explored values of $\lambda_{0,1}$ and $p$ also for deeper
traps, up to $V_0$=25\,$\hbar\omega_0$, and therefore it appears to be
an invariant characteristic of the sinusoidal potential. \par The
result of the numerical integration of Eqs.~(\ref{fdiff}) reported in
Fig.\ref{fig3} reproduces relatively well the subharmonic resonance,
which in the harmonic case would be expected at $\omega_0$. On the
contrary, both experiment and calculation show that the actual
position of the resonance is $\Omega \simeq 0.9\,\omega_0$. It must be
mentioned that the accuracy of these results is restricted by the
finite difference character of Eqs.~(\ref{fdiff}) and by the fact that
some noise sources which have not been included could be resonant at a
nearby frequency. In particular, a possible modulation of the laser
pointing associated to the intensity modulation is expected to be
resonant at $\Omega = \omega_0$ in the harmonic problem
\cite{savard97}, and it could play an analogous role in our sinusoidal
lattice.\par The higher order resonance around 3.5\,$\omega_0$
observed in the experiment is also well reproduced by the calculations
based on first order perturbation theory. Note that a simpler
approximation to the confining potential by a quartic potential
$V_Q(z) = k_2 z^2 + k_4 z^4$ would yield a resonance around
$4\,\omega_0$ and not $3.5\,\omega_0$.  Anyway, it is possible to
understand qualitatively one of the features of this resonance
considering a quartic perturbation of the form $\epsilon(t)V_Q$ to a
harmonic potential. In this case the ratio of the transition rates at
the 2$\omega_0$ and 4$\omega_0$ resonances is set by Eqs.\ref{Rmn:1}
to
\begin{equation}
\vert T(n\pm 2,n)\vert^2/\vert T(n\pm 4,n)\vert^2\propto
\frac{V_0^2}{\omega_0^2}\,.
\end{equation}
This result can qualitatively explain the absence of the
corresponding high order resonance in the {\em radial} excitation
spectrum (see Fig.\ref{fig2}): since the radial trap frequency is a
factor 500 smaller than the axial one, the relative strength of such
radial anharmonic resonance is expected to be suppressed by a factor
(500)$^2$.  In conclusion, high harmonics resonances, which certainly
depend on the actual shape of the anharmonic potential, are expected to
appear only if the spring constant of the trap is large.\par

\begin{figure}
\begin{center}
\leavevmode\epsfxsize=9cm
    \epsfbox{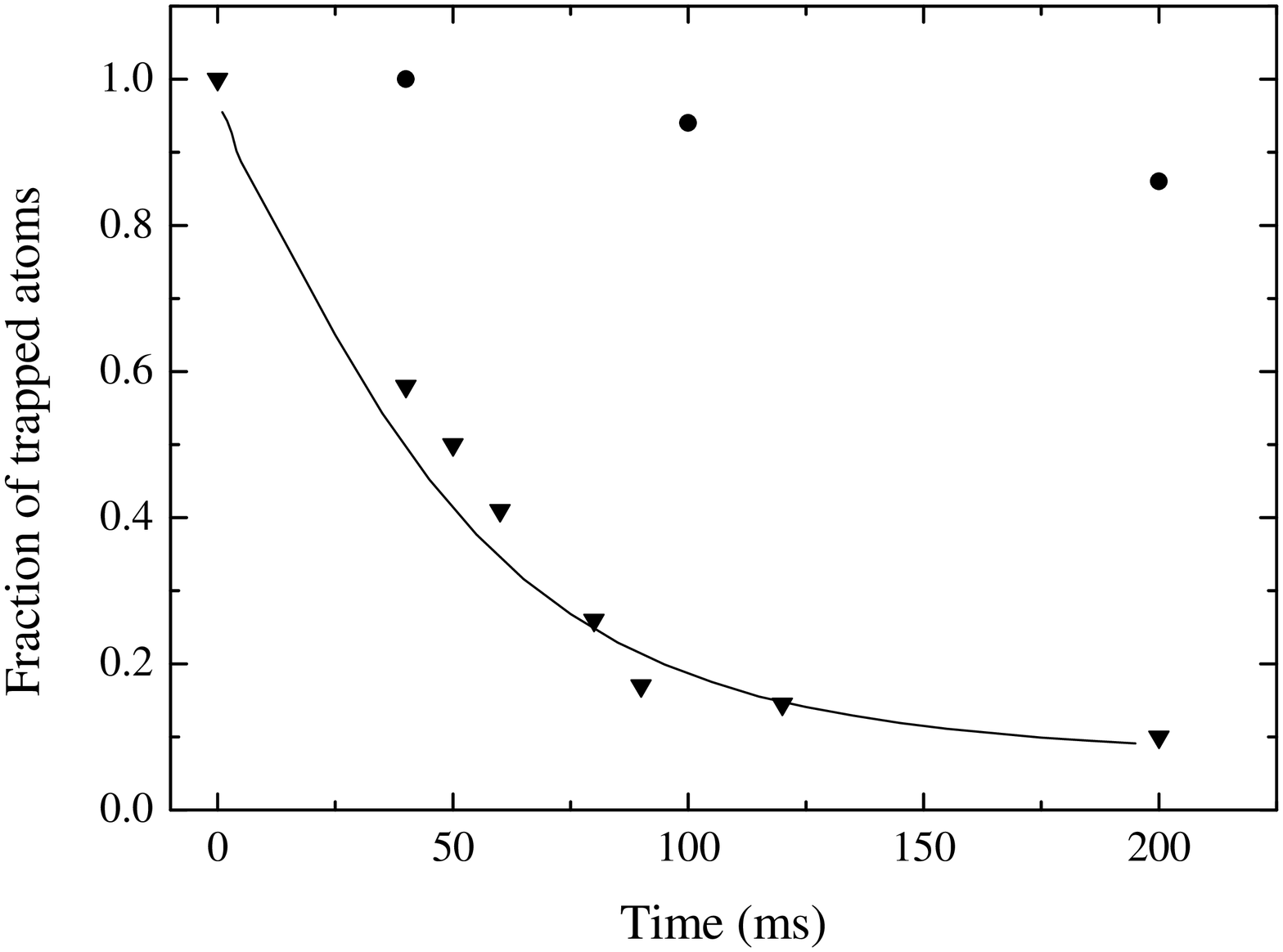}
\end{center}
\caption{ Theoretical (continuous line) and 
  experimental (triangles) results for the evolution of the population
  of trapped atoms at the resonant exciting frequency
  $\Omega=2\omega_0$. The circles show the evolution of the population
  in absence of modulation.}
\label{fig4}
\end{figure}

In Fig.\ref{fig4} theoretical and experimental results for the
evolution of the total population of trapped atoms at the resonant
exciting frequency $\Omega= 2 \omega_0$ are shown.  Although there is
a satisfactory agreement between the model and the experiment, we
notice that experimental data exhibit a different rate for the loss of
atoms before and after 100\,ms. This change is probably due to a
variation of the collision rate as the number of trapped atoms is
modified, which cannot be easily included in the model.  The
comparison of the experimental evolution of the trap population with
and without modulation shows the effectiveness of the excitation
process in emptying the trap on a short time-scale.\par
\begin{figure}
\begin{center}
\leavevmode\epsfxsize=9cm
    \epsfbox{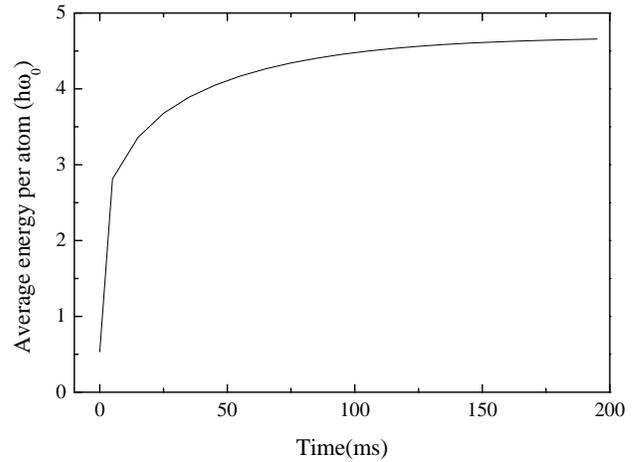}
\end{center}
\caption{Calculated evolution of the average energy of the trapped
  atoms during parametric excitation at $\Omega=2\omega_0$.}
\label{fig5}
\end{figure}

 We have also
simulated the energy growth of the trapped atoms due to the parametric
excitation, which is reported in Fig.\,\ref{fig5}. Our calculations
show nonexponential energy increase in contrast with what expected in
the harmonic approximation, Eq.~(\ref{hexp:1}). The fast energy growth
at short times is related to the depopulation of the lowest levels,
which are resonant with the 2$\omega_0$ parametric source. The
saturation effect observed for longer times is due to the fact that
the resonance condition is not satisfied for the upper levels so that
they do not depopulate easily.

\section{Conclusions.}

We have studied both theoretically and experimentally the time
evolution of the population of atoms trapped in a 1D sinusoidal
optical lattice, following a parametric excitation of the lattice
vibrational mode.  In detail, we have presented a theoretical model
for the excitation in an anharmonic potential, which represents an
extension of the previous harmonic models, and we have applied it to
the actual sinusoidal potential used to trap cold potassium atoms. The
simulation seems to reproduce relatively well the main features of
both the spectrum of trap losses, including the appearance of
resonances beyond 2$\omega_0$, and the time evolution of the total
number of trapped atoms.\par 

By comparing the theoretical predictions and the experimental
observations the usefulness of a parametric excitation procedure to
characterize the spring constant of the trap has been verified.
Although the loss resonances are red-shifted and wider than what
expected in the harmonic case, the lattice harmonic frequency can be
easily extracted from the experimental spectra, to estimate useful
quantities such as the trap depth and spring constant.\par

We have also made emphasis on the need of modeling the broadening of
bands with a negligible natural width in order to reproduce the
observed loss spectrum. In a harmonic model this broadening is not
necessary since the equidistant energy spectrum guarantees that a
single transition energy characterizes the excitation process. We
think that most of the broadening in our specific experiment is due to
the fact that the actual trapping potential is not one-dimensional,
and also to possible fluctuations and inhomogeneities of the lattice.

To conclude, we note that the dynamical analysis we have made can be
easily extended to lattices with larger dimensionality, and also to
other potentials, such as Gaussian potentials, which are
also commonly used for optical trapping.

We acknowledge illuminating discussions with R. Brecha. This work was
supported by the European Community Council (ECC) under the Contracts
HPRI-CT-1999-00111 and HPRN-CT-2000-00125, and by MURST under the PRIN
1999 and PRIN 2000 Programs.

\begin{table}
\begin{tabular}{cccc}
\hline
  r & $E_r$&$E_{r+1}-E_r$ & $\sigma_r$\cr
 \hline
 0&  0.494& 0.976&0.014\cr
 1&  1.470& 0.95&0.015\cr
 2&  2.420 &0.923 &0.019\cr
 3&  3.343&0.897& 0.025\cr
 4&  4.240&0.867& 0.032\cr
 5&  5.107&0.837 &0.042\cr
 6&  5.944&0.802 &0.051\cr
 7&  6.746&0.767 &0.062\cr
 8&  7.513 &0.727 &0.072\cr
 9&  8.240 &0.680 &0.082\cr
 10& 8.920 &0.624& 0.092\cr
 11& 9.544 &0.551 &--\cr
 12& 10.095 & 0.402&--\cr
 13& 10.497 &--& --\cr
\hline
\label{tab1}
\end{tabular}
\caption{ Energy spectrum in units of $\hbar\omega_0$ obtained from the
 diagonalization of the Hamiltonian Eq.~(\ref{h:s}) for 
$V_0$=10.5\,$\hbar\omega_0$ in a harmonic
 basis set with the lowest 40 functions. The third column
 shows the band widths $\sigma_r$, Eqs.~(\ref{S:discrete})and
 (\ref{sigma:r}), used in the numerical simulations reported in Section IV.}
\end{table}

\begin{thebibliography}{99}



\bibitem{savard97} T. A. Savard, K. M. O'Hara, and J. E. Thomas,
{\it Phys. Rev.} A {\bf 56}, R1095 (1997).

\bibitem{friebel98} S. Friebel, C. D'Andrea, J. Walz, M. Weitz, and
  T. W. H\"ansch, Phys. Rev. A {\bf 57}, R20 (1998).

\bibitem {gardiner2000} C. W. Gardiner,  J. Ye, H. C. Nagerl,
 and H. J. Kimble, {\it Phys. Rev. }A {\bf 61}, 045801 (2000).

\bibitem{miller93} J. D. Miller, R. A. Cline, and D. J. Heinzen,
{\it Phys. Rev.} A{\bf 47}, R4567 (1993).

\bibitem{vuletic98} V. Vuletic, C. Chin, A. J. Kerman, and S. Chu,
{\it Phys. Rev. Lett.} {\bf 81}, 5768 (1998).

\bibitem{roati01} G. Roati, W. Jastrzebski, A. Simoni, G. Modugno, and M.
Inguscio, to appear in {\it Phys. Rev.} A; e-print: arXiv physics/0010065.

\bibitem{vuletic99} V. Vuletic, A. J. Kerman, C. Chin, and S. Chu,
{\it Phys. Rev. Lett.} {\bf 82}, 1406 (1999).

\bibitem{landau} L. D. Landau and E. M. Lifshitz, {\it Mechanics}
(Pergamon, Oxford, 1976).

\bibitem{tunnel} We estimate the mean velocity of the atoms along the
  lattice from the energy width of the levels as ${\bar v}=\Delta\omega/2k$.

\bibitem{jauregui01} R. J\'auregui, submitted for publication.

\bibitem{abramowitz}{\it Handbook of Mathematical Functions},
edited by M. Abramowitz and I. A. Stegun (Dover Publications, New
York, 1965).





\end{thebibliography}
\end{document}